\documentclass{sigchi-ext}
\usepackage[T1]{fontenc}
\usepackage{textcomp}
\usepackage[scaled=.92]{helvet} % for proper fonts
\usepackage{graphicx} % for EPS use the graphics package instead
\usepackage{balance}  % for useful for balancing the last columns
\usepackage{booktabs} % for pretty table rules
\usepackage{ccicons}  % for Creative Commons citation icons
\usepackage{ragged2e} % for tighter hyphenation
\usepackage[nolist,nohyperlinks]{acronym}
\usepackage{subfig}
\copyrightinfo{Copyright \textcopyright~2019 for this paper held 
by its author(s). Copying permitted for private and academic 
purposes.} 

\def\plaintitle{Enabling Tangible Interaction through Detection and Augmentation of Everyday Objects} 
\def\plainauthor{Thomas Kosch, Albrecht Schmidt}
\def\plainkeywords{Object Tracking; Computer Vision; In-Situ Assistance; Assistive Systems; Workload-Aware Interfaces}

\title{Enabling Tangible Interaction through Detection and Augmentation of Everyday Objects}

\teaser{
\vspace{-52mm}
{%
\setlength{\fboxsep}{0pt}%
\setlength{\fboxrule}{2pt}%
\fbox{\includegraphics[width=\columnwidth]{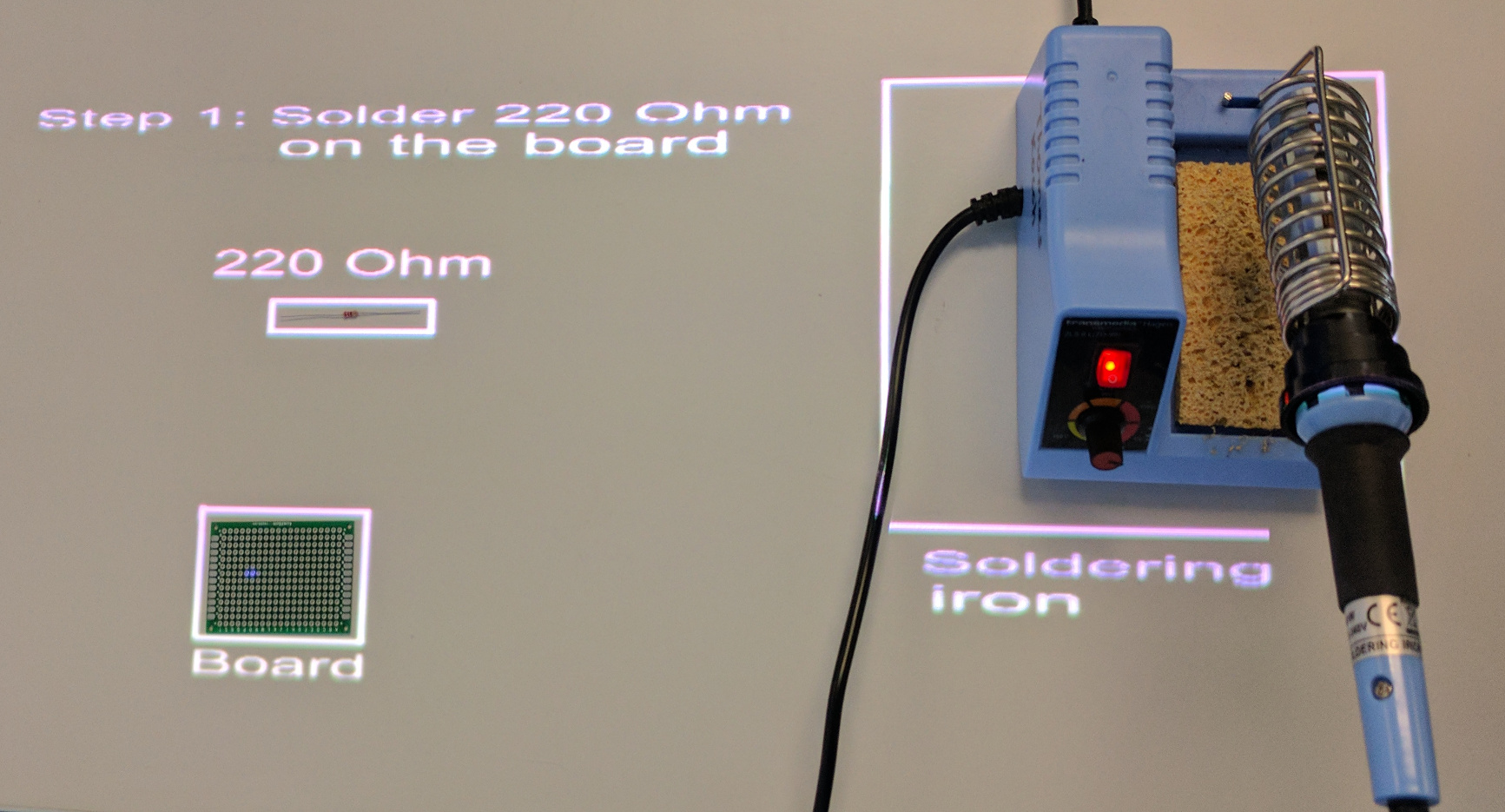}}%
}%
\caption{Augmented workspace in the context of electrical engineering. In-situ projections augment and trace objects as well as track user actions.}
  \label{fig:teaser}
}

\numberofauthors{2}
\author{%
  \alignauthor{%
    \textbf{Thomas Kosch}\\
    \affaddr{LMU Munich} \\
    \affaddr{Munich, Germany} \\
    \email{thomas.kosch@ifi.lmu.de} }\alignauthor{%
    \textbf{Albrecht Schmidt}\\
    \affaddr{LMU Munich}\\
    \affaddr{Munich, Germany}\\
    \email{albrecht.schmidt@ifi.lmu.de} } }

\definecolor{linkColor}{RGB}{6,125,233}
\hypersetup{%
  pdftitle={\plaintitle},
  pdfauthor={\plainauthor},
  pdfkeywords={\plainkeywords},
  bookmarksnumbered,
  pdfstartview={FitH},
  colorlinks,
  citecolor=black,
  filecolor=black,
  linkcolor=black,
  urlcolor=linkColor,
  breaklinks=true,
}

\begin{document}

%% For the camera ready, use the commands provided by the ACM in the Permission Release Form.
%\CopyrightYear{2007}
%\setcopyright{rightsretained}
%\conferenceinfo{WOODSTOCK}{'97 El Paso, Texas USA}
%\isbn{0-12345-67-8/90/01}
%\doi{http://dx.doi.org/10.1145/2858036.2858119}
%% Then override the default copyright message with the \acmcopyright command.
%\copyrightinfo{\acmcopyright}

\maketitle

% Uncomment to disable hyphenation (not recommended)
% https://twitter.com/anjirokhan/status/546046683331973120
\RaggedRight{} 

% Do not change the page size or page settings.
\begin{abstract}
Digital interaction with everyday objects has become popular since the proliferation of camera-based systems that detect and augment objects "just-in-time". Common systems use a vision-based approach to detect objects and display their functionalities to the user. Sensors, such as color and depth cameras, have become inexpensive and allow seamless environmental tracking in mobile as well as stationary settings. However, object detection in different contexts faces challenges as it highly depends on environmental parameters and the conditions of the object itself. In this work, we present three tracking algorithms which we have employed in past research projects to track and recognize objects. We show, how mobile and stationary augmented reality can be used to extend the functionalities of objects. We conclude, how common items can provide user-defined tangible interaction beyond their regular functionality. 

%The deployment of assistive technologies that detect and augment objects "just-in-time" at practical teaching sessions has constantly increased in recent years. The main reason is the ability to provide assistance in industrial and social facilities. Based on the individual learning progress, assistive technologies offer the possibility to augment objects in-situ, thus making it easier for students to grasp the current learning content. In this work, we present three strategies of detecting objects using a camera-based approach and showcase how spatial augmented reality is used to display interaction cues. We summarize and discuss the outcomes of user studies that resulted in evaluating our prototypes. We conclude, how assistive technologies can be used to create smart environments that ubiquitously integrate into learning environments.  

\end{abstract}

\keywords{\plainkeywords}

\category{H.5.m}{Information interfaces and presentation (e.g.,
  HCI)}{Miscellaneous}
  
\section{Introduction}
Augmenting common items with digitized content to extend their functionalities has been the focus of past research in the domain of tangible user interfaces~\cite{Hornecker:2006:GGT:1124772.1124838}. Thereby, objects are tracked by a system that displays visual cues or extends the functionality of the object itself~\cite{Kaltenbrunner:2007:RCF:1226969.1226983}. By rotating, repositioning, or placing objects in defined positions, user-defined actions can be triggered. Thus, common items are augmented by functionalities which they do not implement by themselves.

%Conveying learning content is experiencing a shift from traditional media to interactive learning platforms~\cite{parnell2012idocument}. Providing dynamic information on a dedicated display that augments the environment has shown great potential for delivering learning content based on the individual skills students provide~\cite{Lee2012}. Augmented Reality (AR) has been utilized as a key technology that found its way in vocational schools and universities. 

Two modalities to display such augmented content have emerged. Smart glasses, such as the Microsoft HoloLens\footnote{\url{www.microsoft.com/en-us/hololens} - last access 2019-05-17}, enable mobile use of augmented reality to display additional supporting content~\cite{DANGELMAIER2005371}. Furthermore, in-situ projection systems enable the augmentation of stationary workstations that can be used for practical exercises (see Figure~\ref{fig:teaser}). While smart glasses are preferred in a mobile context, in-situ projections are suitable for stationary settings. While mobile augmentation was preferred during practical physics exercises that required mobility of their students~\cite{Strzys_2018}, industrial use cases~\cite{funk2017working} and social housing organizations~\cite{kosch2019thedigital, kosch2018smart} found stationary settings more suitable. Furthermore, employing object augmentation provides cognitive alleviation, which has the potential to boost overall user performance and productivity~\cite{kosch2017one, kosch2018identifying}.  

Both modalities use camera-based systems to recognize objects and enrich them with additional content. However, seamless object detection and augmentation poses challenges for different use cases. In this work, we present object detection strategies we employed in past research projects to enable object detection and augmentation. We discuss the advantages and disadvantages of different object tracking modalities. Finally, we present how user-defined tangibles from everyday items can be created by augmenting them with in-situ projections. We conclude with challenges that have to be considered when integrating ubiquitous object augmentation.

\begin{figure}
    \centering
    \hspace{-0.75em}
    \subfloat[][]{
        \includegraphics[height=0.24\columnwidth]{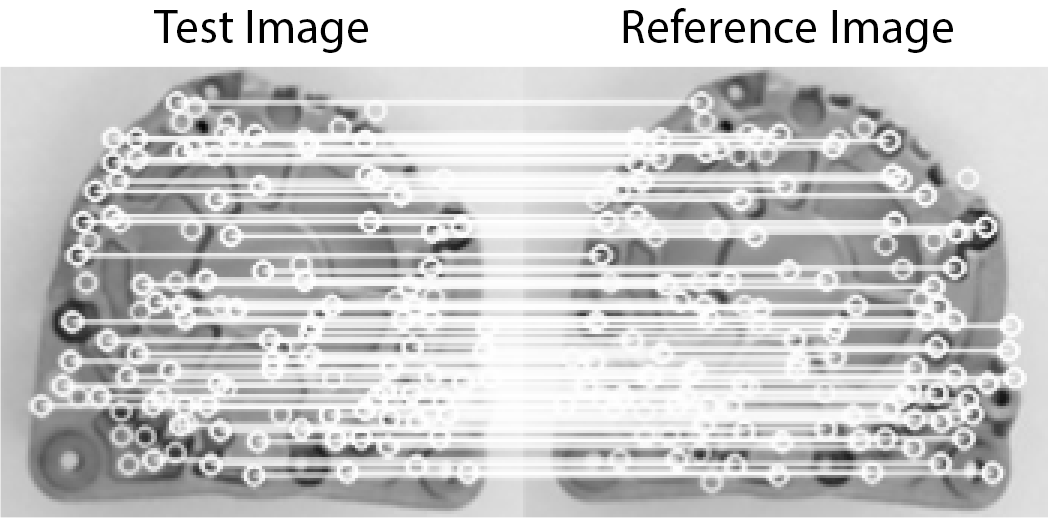}
        \label{fig:lego}}
    \subfloat[][]{
        \includegraphics[height=0.24\columnwidth]{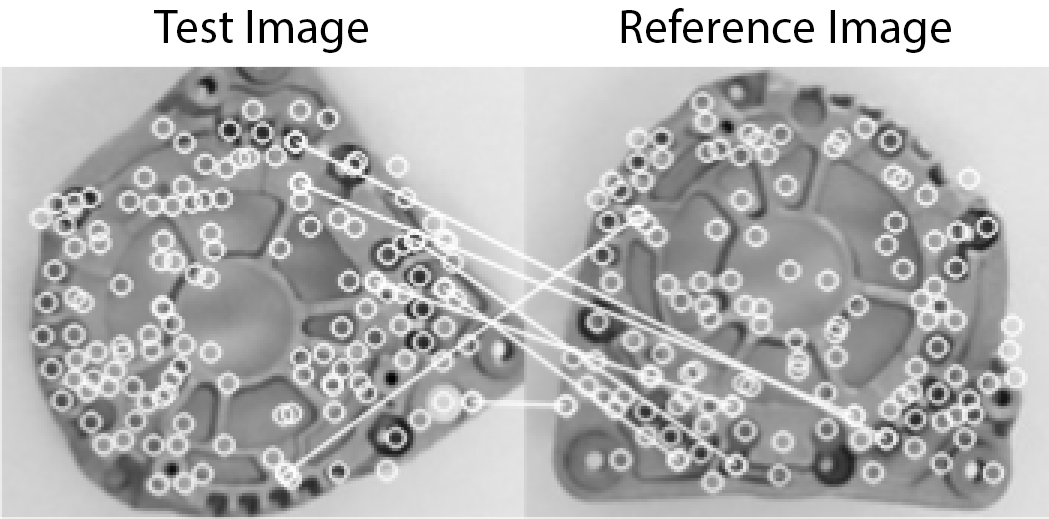}
        \label{fig:box}}
    \caption{Object detection using SURF. The positioned object is compared to a reference image. Feature extraction, such as provided by the SURF algorithm, shows the similarity of the image. \textbf{(a):} Correct positioned image. \textbf{(b):} A rotated object does not guarantee that it will be detected relative to the reference image.}
    \label{fig:surf}
\end{figure}

\section{Object Tracking}
To enable interaction with common items, suitable tracking systems and algorithms need to be employed. In the following, we present three object tracking strategies we have employed in past research.
\subsection{SURF}
The Speeded Up Robust Feature (SURF) algorithm~\cite{5474078} enables to recognize points and areas of "interest" in images. Due to its efficient implementation, it enables the processing of images in real-time. Thereby, the algorithm has been used for object detection by comparing points of interest in a captured image relative to a reference image~\cite{10.1007/11744023_32}. SURF can be employed with inexpensive hardware since it processes color images. However, SURF is not rotation and perspective invariant. This requires objects to be in a similar position that is expected by a system (see Figure~\ref{fig:surf}). 
\subsection{Depth Sensing}

\begin{marginfigure}[1em]
    \begin{minipage}{\marginparwidth}
      \centering
      \includegraphics[width=0.9\marginparwidth]{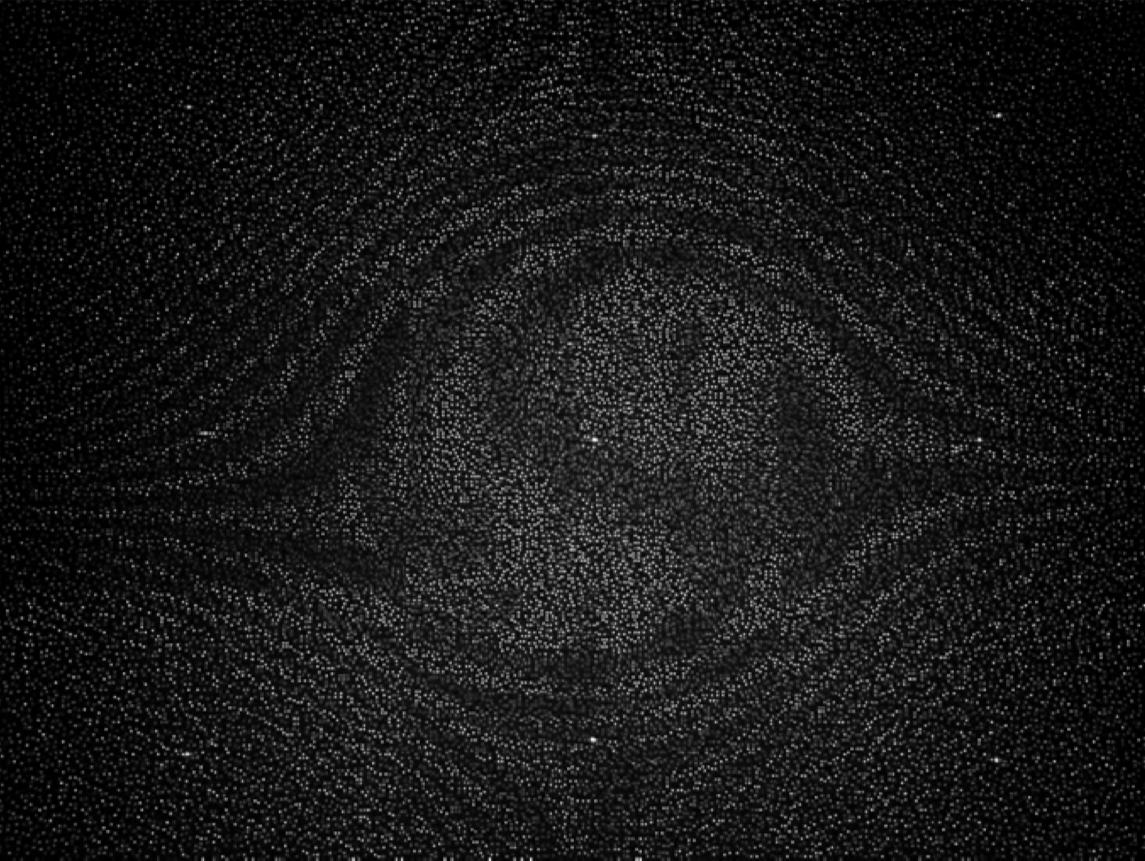}
      \caption{Infrared pattern of a Kinect v1 on a wooden plate \protect\cite{andersen2012kinect}.}
        \label{fig:pattern}
    \end{minipage}
  \end{marginfigure}

A depth sensor, such as the Intel Realsense\footnote{\url{www.intel.com/content/www/us/en/architecture-and-technology/realsense-overview.html} - last access 2019-05-17} or the Kinect v2\footnote{\url{https://developer.microsoft.com/en-us/windows/kinect} - last access 2019-05-17}, provide a 3D representation of objects that they are pointed to. Objects are recognized by analyzing the shape. Thereby, two relevant methods have emerged. The first method uses a projected infrared pattern on a surface (see Figure~\ref{fig:pattern}). Afterward, the depth sensor measures changes in the perspective of the pattern. This enables to detect the distance between infrared waves and allows a reconstruction of the 3D space on a surface~\cite{andersen2012kinect}. 
\begin{figure}
    \centering
    \subfloat[][]{
        \includegraphics[height=0.37\columnwidth]{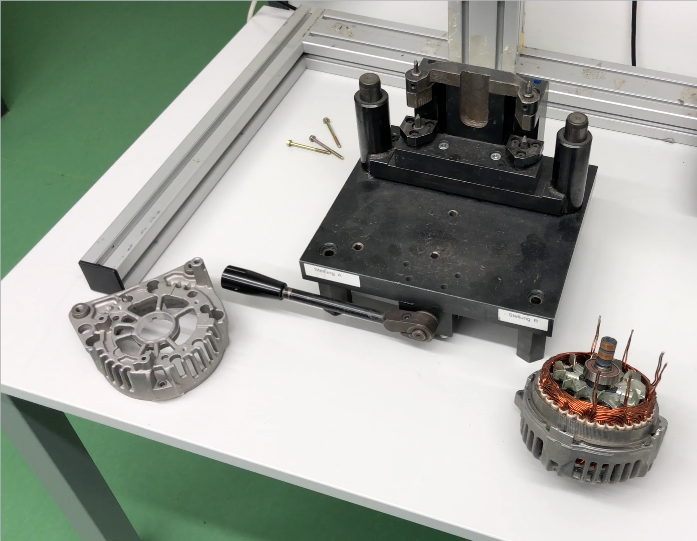}
        \label{fig:lego}}
    \subfloat[][]{
        \includegraphics[height=0.37\columnwidth]{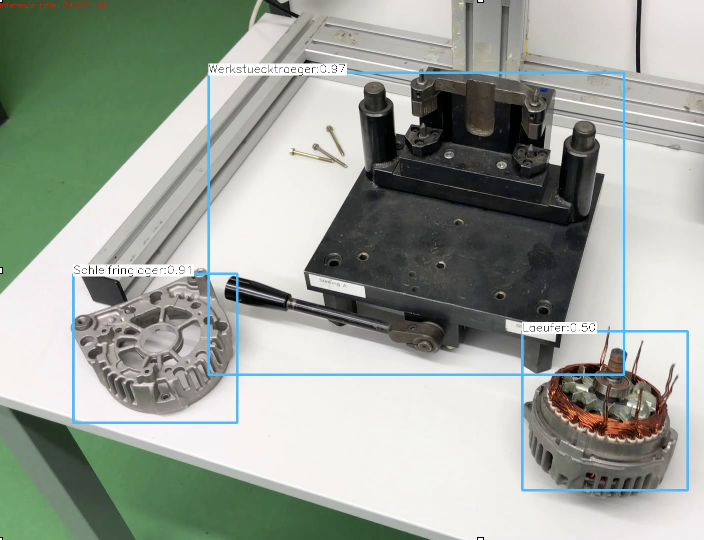}
        \label{fig:box}}
    \caption{Using YOLO to detect objects independent from their position. \textbf{(a):} Test image to evaluate a trained model. \textbf{(b):} Detected objects using YOLO. A blue bounding box denotes the detected objects.}
    \label{fig:yolo}
\end{figure}
The second method uses a Time-of-Flight approach. Thereby, the round trip time of an artificial light (i.e., infrared light) is measured between the sensor and a point on the surface. When the reflection of the light is captured, a 3D representation on of the surface is created~\cite{1384826}.

Depth sensing is insensitive to lighting conditions. However, changes in perspective and rotation of objects may affect the overall detection quality. Thus, depth sensing is suitable for use cases where objects reside in stable positions.

\subsection{You Only Look Once}
The algorithm "You Only Look Once" (YOLO) is a deep learning approach to detect objects regardless of their perspective and position~\cite{Redmon_2016_CVPR} (see Figure~\ref{fig:yolo}). It applies a single neural network on an image that detects features in bounding boxes after clustering their properties. By evaluating those properties, a probability of a correctly detected object is calculated. While YOLO represents a robust real-time method to detect objects regardless of their positioning and perspective, it requires an extensive training set beforehand. Furthermore, training a neural network on a large data set requires time and, depending on the use case, fast computational hardware to speed up the training process. 

\section{Object Augmentation}
Objects can be used as a visual cue for interaction or interaction device itself. In the following, we show implementations of tangible object augmentation we have conducted in the past. 
\subsection{Ambient Augmentation}

\begin{figure}
\hspace{-0.75em}
    \centering
    \subfloat[][]{
        \includegraphics[height=0.32\columnwidth]{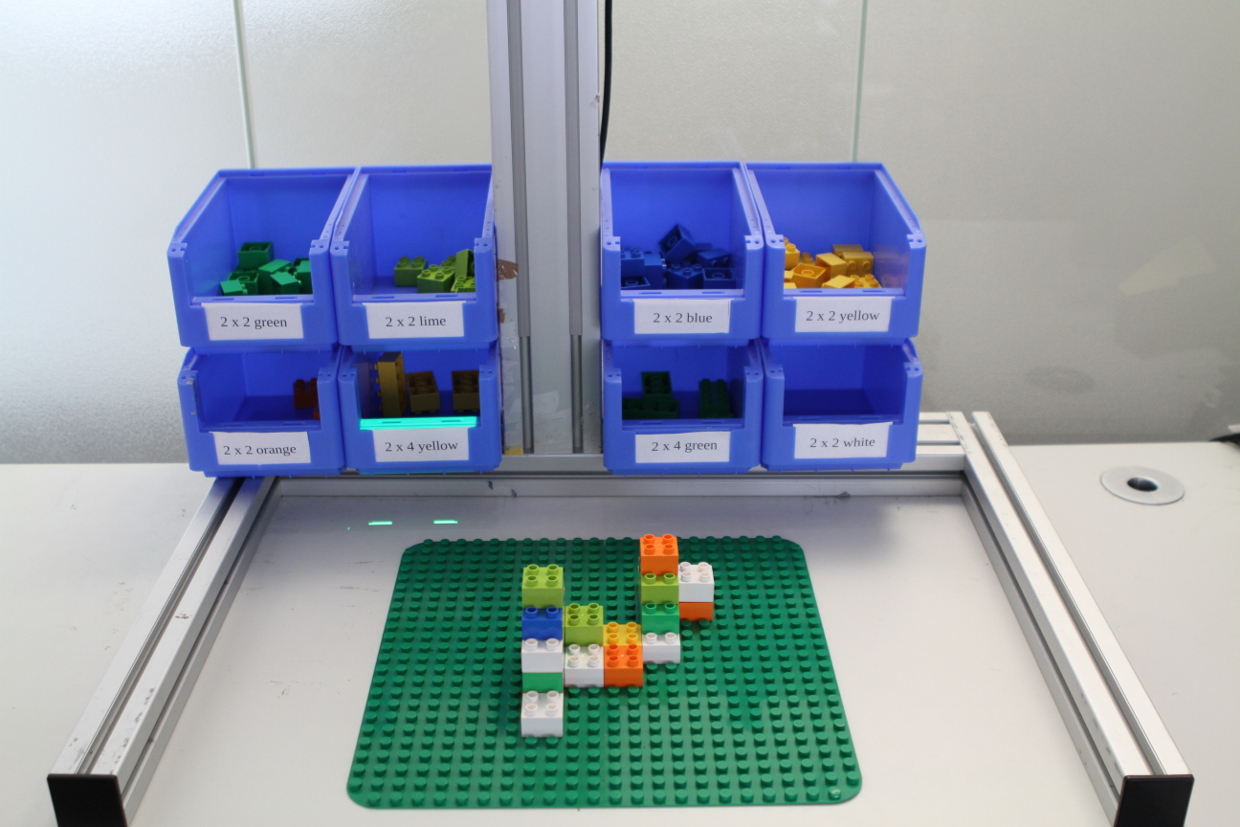}
        \label{fig:lego}}
    \subfloat[][]{
        \includegraphics[height=0.32\columnwidth]{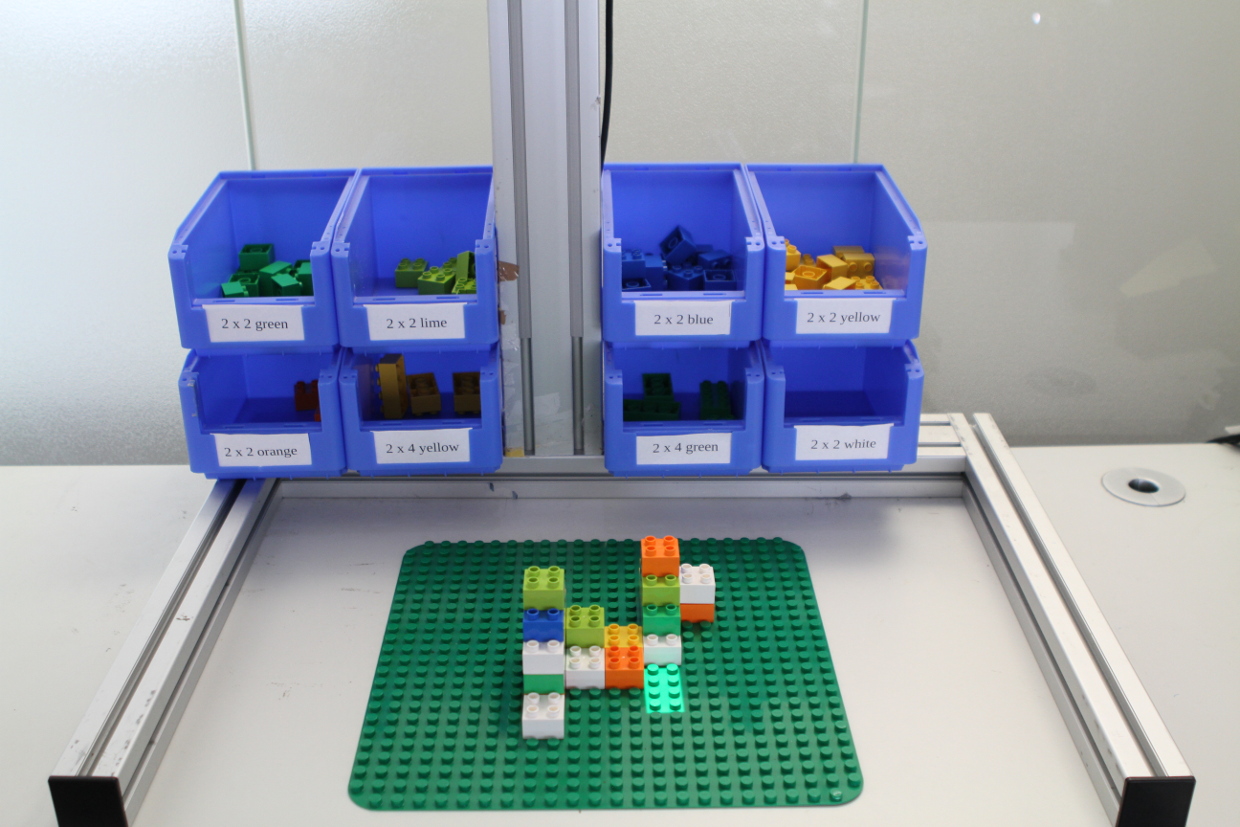}
        \label{fig:box}}
    \caption{Augmenting a workplace using in-situ projections. \textbf{(a):} A detected item selection bin is visually highlighted. \textbf{(b):} A projection on the working area depicts the final position of an assembly part.}
    \label{fig:workspace}
\end{figure}
After recognizing the type of object, cues can be used to implicitly guide the user through a series of actions. Figure~\ref{fig:workspace} shows an augmented workspace that uses in-situ projection as a guide through a series of assembly steps. By detecting the user's action and items on the workspace, in-situ projections are placed on the current relevant bin or final spot for assembly. While boosting the overall performance of workers in industrial environments~\cite{funk2016interactive}, people with dementia and loss in memory benefit from in-situ projections~\cite{kosch2016comparing}.

\subsection{User-defined Tangibles}
\begin{figure}[h!]
\hspace{-0.75em}
    \centering
    \subfloat[][]{
        \includegraphics[height=0.5\columnwidth]{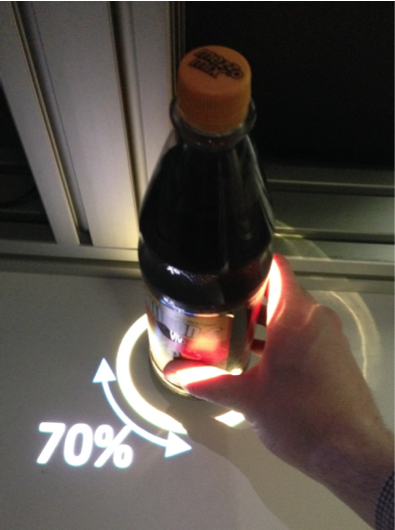}
        \label{fig:knob}}
    \subfloat[][]{
        \includegraphics[height=0.5\columnwidth]{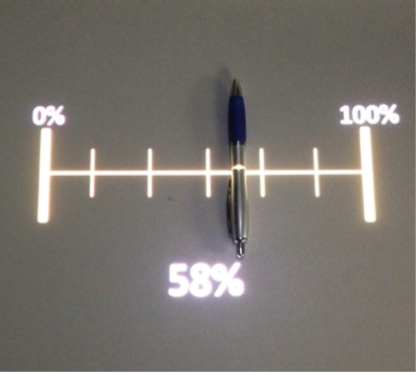}
        \label{fig:slider}}
    \caption{User-defined tangibles that use in-situ projections to provide feedback. \textbf{(a):} Rotating a bottle similar to a knob. \textbf{(b):} Using a pen as a slider \protect\cite{funk2014userdefined}.}
    \label{fig:interaction}
\end{figure}

Regular objects can be registered as user-defined tangible that is made available for interaction~\cite{funk2014userdefined}. For example, rotating (see Figure~\ref{fig:knob}) or positioning (see Figure~\ref{fig:slider}) objects can be used to change the speaker volume. 

After registering the object, a series of options are made available to the user. The user can choose to interact with existing objects or register new objects. Such objects can be everyday items which do not implement a logic. This transforms objects into user-defined tangibles that are already around the user with just-in-time interaction.  

\section{Challenges and Future Work}
Seamless object detection and augmentation in home and workplace settings are prone to certain challenges. In this work, we presented three strategies to detect objects and augment objects. However, choosing the right detection modality depends on the environment as well as on the properties of the object itself. For example, a depth sensor will struggle to detect flat objects as they have scarce 3D properties. While a regular color camera can solve this problem, it is sensitive to the overall environmental illumination. In future work, we want to combine the definition and detection of user-defined tangibles by using an approach that combines color as well as depth images~\cite{5980377}. Thereby, a combination of depth and color data provides an approximation of object type.

Furthermore, privacy and ethical considerations have to be taken into account. By using the presented camera-based approach, public and private spaces are recorded during user interaction. While users can give consent to process the collected data in private settings, public spaces and workplaces are more sensitive to privacy-related issues. In future work, we want to investigate those ethical ramifications. Ultimately, we will investigate design guidelines that explore how a camera-based approach can be conducted while minimally invading the user's privacy.

\section{Conclusion}
In this work, we present three strategies to detect objects which we have employed in past research projects. We outline the advantages and disadvantages of each strategy which we have encountered. We show how object detection and user-defined tangibles can be implemented to provide ambient or explicit interaction. Finally, we discuss challenges that have to be tackled before enabling seamless object tracking in home and work settings.
Since common objects do not implement any logic, we believe that external object augmentation paves the way for ubiquitous tangible interaction at home, public spaces, and workplaces.
\balance{} 
\bibliographystyle{SIGCHI-Reference-Format}
\bibliography{sample}

\end{document}